# Correlated electro-optical and structural study of electrically tunable nanowire quantum dot emitters

Maria Spies,[†] Akhil Ajay,[‡] Eva Monroy,[‡] Bruno Gayral,[‡] and Martien I. den Hertog[†]

[†]Univ. Grenoble-Alpes, CNRS, Institut Néel, 25 av. des Martyrs, 38000 Grenoble, France
[‡]Univ. Grenoble-Alpes, CEA, IRIG-PHELIQS-NPSC, 17 av. des Martyrs, 38000 Grenoble, France

OrcIDs:
Maria Spies: 0000-0002-3570-3422
Akhil Ajay: 0000-0001-5738-5093
Eva Monroy: 0000-0001-5481-3267
Martien I. den Hertog: 0000-0003-0781-9249

**ABSTRACT:** Quantum dots inserted in semiconducting nanowires are a promising platform for the fabrication of single photon devices. However, it is difficult to fully comprehend the electro-optical behaviour of such quantum objects without correlated studies of the structural and optical properties on the same nanowire. In this work, we study the spectral tunability of the emission of a single quantum dot in a GaN nanowire by applying external bias. The nanowires are dispersed and contacted on electron beam transparent $Si_3N_4$ membranes, so that transmission electron microscopy observations, photocurrent and micro-photoluminescence measurements under bias can be performed on the same specimen. The emission from a single dot blue or red shifts when the external electric field compensates or enhances the internal electric field generated by the spontaneous and piezoelectric polarization. A detailed study of two nanowire specimens emitting at 327.5 nm and 307.5 nm shows spectral shifts at rates of 20 and 12 meV/V, respectively. Theoretical calculations facilitated by the modelling of the exact heterostructure provide a good description of the experimental observations. When the bias-induced band bending is strong enough to favor tunneling of the electron in the dot towards the stem or the cap, the spectral shift saturates and additional transitions associated to charged excitons can be observed.

**KEYWORDS:** *Quantum dot, GaN, nanowire, heterostructure, quantum-confined Stark effect, microphotoluminescence*

There is a demand for single photon sources as crucial components for quantum information technologies[1,2] which should open up immense yet unseen possibilities of data treatment. Quantum cryptography, quantum memories or quantum metrology are examples of potential applications.[3,4] Single photon sources can be implemented using two-level emitters such as a semiconductor quantum dot.[5,6] In particular, III-nitride semiconductors offer significant advantages due to their large exciton binding energies and high band offsets, which enable room temperature operation of single photon sources at ultraviolet/visible emission wavelengths.[7,8] Self-assembled GaN quantum dots are generally synthesized following the Stranski-Krastanov (SK) method,[9–11] which takes advantage of the lattice mismatch between different III-nitride compounds to attain three-dimensional (3D) growth. In comparison with SK growth, the synthesis of single quantum dots as axial insertions in nanowires presents the advantage of a larger design flexibility for the dot in terms of size and chemical composition, thanks to the elastic strain relaxation at the nanowire sidewalls.[12] III-nitride quantum dots in nanowires have demonstrated their capability as single photon emitters.[7,8,13]

III-nitride heterostructures present internal electric fields that appear as a result of the discontinuity of polarization (spontaneous and piezoelectric) at the heterointerfaces. The resulting quantum-confined Stark effect (QCSE) has a strong influence on the optical properties, affecting not only the emission wavelength but also the oscillator strength of the radiative transitions. By applying an external bias on a single quantum dot, it is possible to tune the emission spectrally, modify the oscillator strength, and get a better insight on the electronic properties of the dot. Tunable emission from a single SK-grown GaN/AlN quantum dot was demonstrated by Nakaoka et al. in 2006.[14,15] A quantum dot emitting at 3.61 eV ($\approx$ 344 nm) at zero bias blue shifted by 102 meV at +14 V bias[14] (applied along the $c$ crystallographic axis), which means that the spectral shift rate was 7.3 meV per applied Volt. As expected by theory, applying an electric field in the direction of the built-in electric field leads to a red shift in the emission. Likewise applying an external electric field against the built-in electric field compensates the internal electric field, leading to a blue shift of the emission. When bias is applied in-plane,[15] the spectral shift is significantly smaller ($\approx$ 1.1 meV/V for a quantum dot emitting at 366 nm), and the spectra blue shifts symmetrically with respect to the direction of bias.

Müßener et al. demonstrated electrically-tuned single quantum dot emission in a GaN nanowire.[16] Their GaN quantum dot consists of an $Al_{0.3}Ga_{0.7}N/GaN/Al_{0.3}Ga_{0.7}N$ (50 nm / 1.7 nm / 50 nm) insertion sandwiched by 16-nm-thick AlN current blocking layers. The emission of the GaN dot peaked at 3.56 eV ($\approx$ 348 nm) at zero bias, and was shown to shift by 1.3 meV/V in the range of −15 V to +25 V. It was also demonstrated that, in the case of a superlattice of quantum dots embedded in a nanowire, bias can lead to a transition from direct excitons (electron and hole in the same quantum dot) to indirect excitons (electron and hole in different dots).[17] In this case, the direct transition shifted at a rate of 5.3 meV/V for quantum dots emitting around 400 nm at zero bias.

Correlation between the opto-electrical properties and the structure of an individual heterostructured nanowire is crucial to get a complete insight on the electronic behavior and optimize device design. Modern fabrication techniques allow the identification of a single nanowire, its optical and electrical characterization at low temperature, and its observation with scanning transmission electron microscopy (STEM).[18–23] Therefore, it is possible to correlate the spectral features of a single nano-object to its structural features. For example, it has been shown by TEM correlated with transport or optical experiments that in $In_2Se_3$ NWs the conductive behavior (metallic or semiconductor) depends on the crystallographic growth direction of the NWs[22] and in InP NWs that the photoluminescence of single crystalline NWs was very different from the PL of NWs containing twin defects[23], while no strong effect of polytypism was observed in Ga(N)P NWs[24].

In the present paper, we study the effect of external bias on GaN/AlGaN quantum dots contained in GaN nanowires. The design of the structure targets the maximization of the sensitivity to bias, while keeping the emission energy above the GaN bandgap. The downscaling of the heterostructure increases sensitivity to growth parameters, so that there is a certain dispersion in material properties in the nanowire ensemble. By measuring bias-dependent photoluminescence (PL), photocurrent, and STEM on the same specimen, it is possible to correlate the optoelectronic and structural properties of a single dot in a nanowire, undeterred by the statistical variations from wire to wire.

## EXPERIMENTAL RESULTS

**Design, structural characterization and modelling.** The specimens under study were GaN nanowires incorporating an Al(Ga)N/GaN/Al(Ga)N insertion. Nominally, the heterostructure consisted of a 1-nm-thick GaN quantum dot embedded between two 9-nm-thick AlN barriers. The heterostructure was surrounded by segments of undoped GaN (each 130 nm long), whereas the ends of the nanowires were doped at $8\times10^{17}$ cm$^{-3}$ with Ge, to facilitate ohmic contacts. A schematic of the samples is shown in figure 1(a), and the band diagram of the nominal structure obtained from one-dimensional (1D) calculations using the nextnano$^3$ software is displayed in figure 1(b).

The thickness of the quantum dot was chosen so that its emission is energetically higher than the GaN band gap. This choice reduces the sensitivity of the emission wavelength to the electric field, but it allows the unambiguous identification of the quantum dot emission, located at higher energies than any emission from the GaN nanowire stem or cap sections. Regarding the AlN barriers, thicker AlN sections are less sensitive to interface phenomena and help to reduce the leakage current, but they lead to a larger active region, hence more voltage is required to vary the electric field applied on the quantum dot. Furthermore, spontaneous and piezoelectric polarization lead to accumulation of electrons and holes at the heterointerfaces with the GaN stem and cap (see figure 1(b)), which reduces the sensitivity to the external electric field and can favor tunneling transport between the conduction band of the cap and the valence band of the stem, as previously reported in GaN/AlN heterostructures.[25,26] With these considerations in mind, a nominal thickness of 9 nm of AlN for the barriers was considered thick enough to ensure low leakage current. High angle annular dark field (HAADF) STEM images of the two contacted single nanowires analyzed in this paper (NW1 and NW2) are shown in figures 1(c) and (d), respectively, where red squares mark the heterostructures.

Let us first concentrate on the structure of NW1. A zoomed micrograph of the area within the red square in figure 1(c) is displayed in figure 2(a). The GaN dot height is estimated to be 1.2 nm ± 0.5 nm (evaluated by measuring the full width at half maximum of the bright contrast in the HAADF STEM image corresponding to the GaN quantum dot), and the barrier thicknesses are around 9.3 nm (bottom barrier) and 7.3 nm (top barrier). An AlN shell envelops the GaN dot. The bottom barrier presents a large dome-shaped diffusion region of AlGaN with an Al mole fraction around 50%, which extends up to the GaN quantum dot. This composition is estimated from the HAADF contrast in the lower barrier region being approximately halfway between the intensity levels for pure GaN (stem and cap) and AlN (top barrier). Such a diffusion phenomenon has been reported previously and is explained as an out-diffusion of Ga from the stem due to the significant lattice mismatch and high growth temperature.[27] After the growth of the heterostructure, a strain-induced narrowing of the nanowire diameter is observed. With this structural information, the heterostructure in NW1 was modelled in 3D as described in figure 2(b), using nextnano³. The obtained calculation of the internal electric field along the growth axis, taking into account the elastic relaxation of the structure, is presented in figure 2(c). The electric field in the center of the GaN dot is around 4.8 MV/cm.

In the case of NW2, the micrograph in figure 2(d) shows a stronger out-diffusion of Ga, which affects also the top barrier. The GaN dot is thinner, with a thickness around 0.8 nm, and there is no trace of AlN shell around the dot. Modeling the structure as described in figure 2(e) yields an axial electric field of 3 MV/cm in the GaN dot, with the field distribution illustrated in figure 2(f).

**Electrical tuning of the quantum dot emission.** Before recording the photoluminescence (PL) spectra, the current-voltage (I-V) characteristics of the nanowires were recorded in the dark and under illumination with the ultraviolet laser used for micro-photoluminesce (µPL) excitation. Figure 3(a) presents the corresponding I-V characteristics of NW1. The dark current stays below 30 pA in the −4 V to +4 V range. A similar current is obtained when measuring an I-V curve between two contacts on the membrane chip which are not connected by either a metal line or a nanowire. Thus, the measured current is assigned to a leakage current path through the membrane chip. Under ultraviolet illumination, however, NW1 becomes rectifying. Starting around 1.5 V, the current increases, and it reaches almost 2 nA at 2 V.

The spectrum obtained by µPL of NW1 at zero bias shows an emission line from the GaN quantum dot, as illustrated in figure 3(b) (magenta line), which peaks at 327.5 nm. For negative(positive) bias, the emission red(blue) shifts as a result of the enhancement(compensation) of the internal electric field in the dot. The shift occurs at a rate of 20 meV/V in the range from −2 V to +4 V bias, which means 3-10 times higher sensitivity to bias than in previous reports, even if our nanowire is emitting at shorter wavelength (327.5 nm at zero bias, to be compared with 344 nm in ref. [14], 348 nm in ref. [16], 366 nm in ref. [15], or even 400 nm in ref. [17]). Note that the spectra in figure 3(b) are not normalized and are hence comparable in intensity. The PL intensity and linewidth remain approximately stable when applying bias, dropping only for $V_B > 2$ V. This drop correlates with the increase of the photocurrent, i.e. the applied electric field separates the photogenerated electrons and holes hence decreasing the radiative recombination probability.

For comparison, the behavior of NW2 under optical excitation and bias is depicted in figure 3(c-d). Let us remember that this sample contains a smaller dot and barriers with higher Ga content in the barriers, which explains the emission at shorter wavelengths (peak at 307.5 nm at zero bias) and the higher photocurrent (almost double) in comparison with NW1. The red(blue) shift with negative(positive) bias is consistent with figure 3(b). Under negative bias, the spectral shift takes place at a rate of 12 meV/V. This is consistent with the expectation of a lower sensitivity to the electric field in dots emitting at shorter wavelengths. As a particularity, for bias between +0.6 V and +2.8 V, the spectra present two peaks which keep blue shifting with increasing positive bias, i.e. increasingly compensated internal electric field. These features are stable and reproducible after warming up and cooling the nanowire down again. One of the peaks disappears for bias higher than +2.8 V and the second peak also eventually disappears around +4 V bias.

**DISCUSSION**

For the interpretation of these experimental results, figure 4 compares the emission wavelengths with the results of theoretical calculations obtained as described in the Methods section, taking into account the actual nanowire geometry presented in figure 2.

If we focus on NW1, the simulations reproduce with good precision the emission wavelength at zero bias and the evolution of the luminescence under positive bias [see data outlined with a red square in figures 4(a) and (b)]. From these data, we can extract the location of the voltage drop along the nanowire. Comparing figures 4(a) and (b), an applied bias of 4 V corresponds to an applied electric field in the dot of 2.2 MV/cm, which would imply that the voltage drops in a total length of 18 nm, which is approximately the length of the AlGaN/GaN/AlGaN insertion. We can therefore deduce that the resistance of the GaN stem and cap is negligible in comparison with the AlGaN sections.

Under negative bias ($V_A < -2$ V) the experimental peak emission shows a saturation which is not observed in the simulated graph. To understand this deviation, we must look at the calculated band diagrams and electron wavefunctions in the dot, displayed in figure 5(a-c). Using the band diagram at zero bias as a reference [figure 5(b)], positive bias results in partial compensation of the internal electric field [figure 5(c)]. As a result, the wavefunctions of the electron and the hole in the dot get geometrically closer but the transition energy increases. However, under negative bias the electric field in the dot is enhanced [figure 5(a)]. The separation of the electron and the hole increases and the wavefunction of the electron extends now through the lower barrier towards the stem. Furthermore, the electric field in the lower barrier, initially pointing towards the nanowire stem [see figure 5 (b)] is inverted under reverse bias, which favors the electron tunneling from the dot to the stem. This phenomenon is highly sensitive to the Al mole faction of the lower barrier, which is strongly perturbed by the strain-induced out-diffusion of Ga [see figure 2(a)], which explains the deviation between the experimental PL measurements and the simulations in this bias range.

Let us now turn our attention to NW2 in figure 4(c-d). Experiments and simulations present a particularly good match for zero and negative bias. In this case, –4 V bias corresponds to an applied electric field in the dot of 1.9 MV/cm, which implies that the voltage drops along 21 nm, i.e. it is further confirmed that the resistance of the stem and cap is negligible in comparison to the AlGaN sections. Looking at the band diagrams in figures 5(d-f) to understand the nanowire behavior, we find significant differences with NW1. At zero bias [figure 5(e)] the electric field in the dot is smaller for NW2, since the dot itself is geometrically smaller along the nanowire axis. On the contrary, the electric field in the lower barrier is higher in NW2 due to the smaller Al mole fraction in the upper barrier in comparison to NW1 [see HAADF-STEM image in figure 2(d), which displays larger Ga out-diffusion in comparison to NW1 in figure 2(a)]. As a result, the negative bias required to invert the sense of the electric field in the lower barrier and extract the electron from the dot is higher than in NW1. Thus, in figure 5(d), for −1.5 MV/cm applied field, the electron wavefunction remains located in the dot. On the contrary, the presence of Ga in the upper barrier favors the escape of the electron towards the cap under positive bias. The difficulty to confine the electron in the dot might be associated to the observation of the second emission peak. The electron-hole Coulombic interaction is not considered by the calculations, and hole tunneling from the stem might help to stabilize the exciton in the dot. The second emission line is hence assigned to a charged state of the exciton, which seems reasonable in view of the energy separation between the two peaks.

The bias sensitivity of NW1 and NW2 lie almost one order of magnitude above those observed by Müßener et al.[16] in a dot-in-wire structure. The higher applied voltages required in their case can be explained by the thickness of their AlGaN+AlN barriers. In the work by Müßener et al., the total thickness of the heterostructures is ≈134 nm, significantly larger than in our design, where the heterostructure is only ≈20 nm thick.

**CONCLUSION**

To conclude, we could spatially isolate single nanowires to perform several spectroscopic (bias-dependent photoluminescence, photocurrent) and structural (STEM) measurements on the same GaN nanowire containing a ≈ 1-nm-thick Al(Ga)N/GaN quantum dot. From these correlated measurements, we could model the evolution of the photoluminescence with bias, taking into account the measured structural parameters of each dot. A systematic blue(red) shift is observed in µPL measurements for the application of an external electric field

compensating(enhancing) the polarization-related internal electric field within the quantum dot structure. Spectral shifts of 12 and 20 meV/V are observed for quantum dots emitting at 307.5 and 327.5 nm, respectively. Three-dimensional modeling of the electronic structure taking the STEM-measured morphology into account allow estimating the internal electric field in the dots (around 3 and 4.8 MV/cm in the two nanowires under study, respectively) and predicting its variation with bias. A deviation from the theoretical trend is observed when the bias voltage is high enough to favor tunneling of the electron in the quantum dot towards the stem or the cap. In such a situation, the spectral shifts saturate and additional transitions associated to charged excitons can be observed.

## METHODS

Self-assembled (000-1)-oriented[20] GaN nanowires were synthesized by plasma-assisted molecular beam epitaxy (PAMBE) on Si(111) substrates. The growth was initiated by deposition of an AlN buffer layer using the two-step growth procedure described in ref. [28]. Then, GaN nanowires were deposited under nitrogen rich conditions (Ga/N flux ratio = 0.25), at a growth rate of ≈ 300 nm/h and a substrate temperature of $T_S$ = 810°C. The AlN sections that define the quantum dot insertion were grown stoichiometrically. The nanowires exhibited a total length of 3.2 µm and diameters around 60 nm. To ensure the homogeneous height of the nanowires before the growth of the heterostructure,[29] the quantum dot insertion is asymmetrically located in the nanowire, 2.2 µm away from the Si substrate.

The as-grown nanowire ensemble was sonicated in isopropanol. The resulting solution was dispersed on silicon chips containing 40-nm-thick 100×100 µm² wide $Si_3N_4$ membranes compatible with transmission electron microscopy experiments. These membrane chips were fabricated from 400-µm-thick n$^{++}$-Si(100) wafers. On each side of the wafers, a bilayer of $SiO_2/Si_3N_4$ (200 nm/ 40 nm) was deposited. Membrane windows and cleave lines were opened using laser lithography, reactive ion etching and a KOH bath. Through another lithography step metallic markers and large contact pads were defined both on and around the membranes. The dispersed nanowires were contacted with electron beam lithography and metal evaporation of Ti/Al (5 nm/120 nm).

Theoretical calculations of the band diagram were performed using the commercial 8×8 k·p band Schrödinger-Poisson equation solver nextnano[30] with the material parameters described in ref. [30]. Three-dimensional (3D) simulations were carried out in order to obtain the strain state and the band diagram. The nanowire was modeled as a hexahedral prism consisting of a 150-nm-long n-type GaN segment followed by 100 nm of undoped GaN, the active heterostructure, 100 nm of undoped GaN and 50 nm of n-type GaN. The dimensions and composition of the heterostructure were chosen as a function of the results of structural characterization of the precise nanowire under study.

The result of 3D calculations was used as an input for 1D calculations of the wire under external bias. In such calculations, the structure presented the same geometry and material composition along the growth axis, but the internal electric field at zero bias was defined to be the result of the 3D calculations. The quantum confined electron levels in the dot were then obtained as a function of an externally applied electric field, using the 8-band k·p model.

Structural information on the nanowire heterostructure was gathered with HAADF STEM using a probe-corrected FEI TITAN Themis working at 80 and 200 kV. A DENSSolutions 6 contact double tilt holder was employed.

I-V characteristics were measured using an Agilent 4155C semiconductor parameter analyzer. In general, the bias was chosen to keep the maximum photocurrent lower than 1 μA to prevent device failure.

Microphotoluminescence (μPL) measurements were carried out using a frequency-doubled solid-state laser emitting at 244 nm, with excitation power up to 1 mW. The laser was focused into a spot with a diameter around 1 µm using a refractive microscope objective (numerical aperture NA=0.4). A Jobin-Yvon tri-axe 550 monochromator and a liquid-nitrogen-cooled ultraviolet-enhanced charge-coupled device (CCD) were used for spectral dispersion and detection. The sample was mounted on the cold finger of an Oxford He-flow cryostat. The visualization of the sample was performed using an LED emitting at 365 nm that illuminates the sample and an ultraviolet-enhanced camera (pco.ultraviolet, 190–1100 nm, 1392×1040 pixel$^2$). The optimization of the alignment of the laser on the nanowire was performed by maximizing the photocurrent. Furthermore, a 300/80 nm single-bandpass filter (BrightLine FF01-300/80-25) was used to diminish the near-band-gap GaN emission line from the nanowire cap/stem. The nanowire was biased using a Keithley 2450 sourcemeter.

## ACKNOWLEDGEMENTS

Financial support from the ANR-UVLASE (ANR-18-CE24-0014) project and the AGIR 2016 Pole PEM funding proposed by Grenoble Alpes University (UGA) for the CoPToN project is acknowledged. A.A. acknowledges financial support from the French National Research Agency via the GaNEX program (ANR-11-LABX-0014). We benefitted from the access to the Nano characterization platform (PFNC) in CEA Minatec Grenoble in collaboration with the IRIG/LEMMA group. Membrane production and nanowire contacting have been carried out at the NanoFab cleanroom of Institut Néel, Grenoble. Thanks are due to Bruno Fernandez and Jean-François Motte for cleanroom support, as well as to Yann Genuist, Yoann Curé and Fabien Jourdan for PAMBE support, and Fabrice Donatini and Jonas Lähnemann for fruitful discussions. This project has received funding from the European Research Council (ERC) under the European Union's Horizon 2020 research and innovation programme "e-See" (Grant Agreement 758385).

# REFERENCES

(1) Claudon, J.; Bleuse, J.; Malik, N. S.; Bazin, M.; Jaffrennou, P.; Gregersen, N.; Sauvan, C.; Lalanne, P.; Gérard, J.-M. A Highly Efficient Single-Photon Source Based on a Quantum Dot in a Photonic Nanowire. *Nat. Photonics* **2010**, *4* (3), 174–177. https://doi.org/10.1038/nphoton.2009.287.

(2) Aharonovich, I.; Englund, D.; Toth, M. Solid-State Single-Photon Emitters. *Nat. Photonics* **2016**, *10* (10), 631–641. https://doi.org/10.1038/nphoton.2016.186.

(3) Beveratos, A.; Brouri, R.; Gacoin, T.; Villing, A.; Poizat, J.-P.; Grangier, P. Single Photon Quantum Cryptography. *Phys. Rev. Lett.* **2002**, *89* (18), 187901. https://doi.org/10.1103/PhysRevLett.89.187901.

(4) Knill, E.; Laflamme, R.; Milburn, G. J. A Scheme for Efficient Quantum Computation with Linear Optics. *Nature* **2001**, *409* (6816), 46–52. https://doi.org/10.1038/35051009.

(5) Shields, A. J. Semiconductor Quantum Light Sources. *Nat. Photonics* **2007**, *1*, 215.

(6) Mäntynen, H.; Anttu, N.; Sun, Z.; Lipsanen, H. Single-Photon Sources with Quantum Dots in III–V Nanowires. *Nanophotonics* **2019**, *8* (5), 747–769. https://doi.org/10.1515/nanoph-2019-0007.

(7) Deshpande, S.; Heo, J.; Das, A.; Bhattacharya, P. Electrically Driven Polarized Single-Photon Emission from an InGaN Quantum Dot in a GaN Nanowire. *Nat. Commun.* **2013**, *4*, 1675. https://doi.org/10.1038/ncomms2691.

(8) Holmes, M. J.; Choi, K.; Kako, S.; Arita, M.; Arakawa, Y. Room-Temperature Triggered Single Photon Emission from a III-Nitride Site-Controlled Nanowire Quantum Dot. *Nano Lett.* **2014**, *14* (2), 982–986. https://doi.org/10.1021/nl404400d.

(9) Kako, S.; Hoshino, K.; Iwamoto, S.; Ishida, S.; Arakawa, Y. Exciton and Biexciton Luminescence from Single Hexagonal GaN/AlN Self-Assembled Quantum Dots. *Appl. Phys. Lett.* **2004**, *85* (1), 64–66. https://doi.org/10.1063/1.1769586.

(10) Bardoux, R.; Guillet, T.; Lefebvre, P.; Taliercio, T.; Bretagnon, T.; Rousset, S.; Gil, B.; Semond, F. Photoluminescence of Single GaN ∕ AlN Hexagonal Quantum Dots on Si ( 111 ) : Spectral Diffusion Effects. *Phys. Rev. B* **2006**, *74* (19), 195319. https://doi.org/10.1103/PhysRevB.74.195319.

(11) Rol, F.; Founta, S.; Mariette, H.; Daudin, B.; Dang, L. S.; Bleuse, J.; Peyrade, D.; Gérard, J.-M.; Gayral, B. Probing Exciton Localization in Nonpolar Ga N ∕ Al N Quantum Dots by Single-Dot Optical Spectroscopy. *Phys. Rev. B* **2007**, *75* (12), 125306. https://doi.org/10.1103/PhysRevB.75.125306.

(12) Gudiksen, M. S.; Lauhon, L. J.; Wang, J.; Smith, D. C.; Lieber, C. M. Growth of Nanowire Superlattice Structures for Nanoscale Photonics and Electronics. *Nature* **2002**, *415* (6872), 617–620. https://doi.org/10.1038/415617a.

(13) Kako, S.; Santori, C.; Hoshino, K.; Götzinger, S.; Yamamoto, Y.; Arakawa, Y. A Gallium Nitride Single-Photon Source Operating at 200 K. *Nat. Mater.* **2006**, *5* (11), 887–892. https://doi.org/10.1038/nmat1763.

(14) Nakaoka, T.; Kako, S.; Arakawa, Y. Quantum Confined Stark Effect in Single Self-Assembled GaN/AlN Quantum Dots. *Phys. E Low-Dimens. Syst. Nanostructures* **2006**, *32* (1–2), 148–151. https://doi.org/10.1016/j.physe.2005.12.028.

(15) Nakaoka, T.; Kako, S.; Arakawa, Y. Unconventional Quantum-Confined Stark Effect in a Single GaN Quantum Dot. *Phys. Rev. B* **2006**, *73* (12), 121305. https://doi.org/10.1103/PhysRevB.73.121305.

(16) Müßener, J.; Teubert, J.; Hille, P.; Schäfer, M.; Schörmann, J.; de la Mata, M.; Arbiol, J.; Eickhoff, M. Probing the Internal Electric Field in GaN/AlGaN Nanowire Heterostructures. *Nano Lett.* **2014**, *14* (9), 5118–5122. https://doi.org/10.1021/nl501845m.

(17) Müßener, J.; Hille, P.; Grieb, T.; Schörmann, J.; Teubert, J.; Monroy, E.; Rosenauer, A.; Eickhoff, M. Bias-Controlled Optical Transitions in GaN/AlN Nanowire Heterostructures. *ACS Nano* **2017**, *11* (9), 8758–8767. https://doi.org/10.1021/acsnano.7b02419.

(18) Ahtapodov, L.; Todorovic, J.; Olk, P.; Mjåland, T.; Slåttnes, P.; Dheeraj, D. L.; van Helvoort, A. T. J.; Fimland, B.-O.; Weman, H. A Story Told by a Single Nanowire: Optical Properties of Wurtzite GaAs. *Nano Lett.* **2012**, *12* (12), 6090–6095. https://doi.org/10.1021/nl3025714.

(19) Zagonel, L. F.; Mazzucco, S.; Tencé, M.; March, K.; Bernard, R.; Laslier, B.; Jacopin, G.; Tchernycheva, M.; Rigutti, L.; Julien, F. H.; et al. Nanometer Scale Spectral Imaging of Quantum Emitters in Nanowires and Its Correlation to Their Atomically Resolved Structure. *Nano Lett.* **2011**, *11* (2), 568–573. https://doi.org/10.1021/nl103549t.

(20) den Hertog, M. I.; González-Posada, F.; Songmuang, R.; Rouviere, J. L.; Fournier, T.; Fernandez, B.; Monroy, E. Correlation of Polarity and Crystal Structure with Optoelectronic and Transport Properties of GaN/AlN/GaN Nanowire Sensors. *Nano Lett.* **2012**, *12* (11), 5691–5696. https://doi.org/10.1021/nl302890f.

(21) den Hertog, M.; Donatini, F.; McLeod, R.; Monroy, E.; Sartel, C.; Sallet, V.; Pernot, J. In Situ Biasing and Off-Axis Electron Holography of a ZnO Nanowire. *Nanotechnology* **2018**, *29* (2), 025710. https://doi.org/10.1088/1361-6528/aa923c.

(22) Peng, H.; Xie, C.; Schoen, D. T.; Cui, Y. Large Anisotropy of Electrical Properties in Layer-Structured $In_2Se_3$ Nanowires. *Nano Lett.* **2008**, *8* (5), 1511–1516. https://doi.org/10.1021/nl080524d.

(23) Bao, J.; Bell, D. C.; Capasso, F.; Wagner, J. B.; Mårtensson, T.; Trägårdh, J.; Samuelson, L. Optical Properties of Rotationally Twinned InP Nanowire Heterostructures. *Nano Lett.* **2008**, *8* (3), 836–841. https://doi.org/10.1021/nl072921e.

(24) Dobrovolsky, A.; Persson, P. O. Å.; Sukrittanon, S.; Kuang, Y.; Tu, C. W.; Chen, W. M.; Buyanova, I. A. Effects of Polytypism on Optical Properties and Band Structure of Individual Ga(N)P Nanowires from Correlative Spatially Resolved Structural and Optical Studies. *Nano Lett.* **2015**, *15* (6), 4052–4058. https://doi.org/10.1021/acs.nanolett.5b01054.

(25) Leconte, S.; Guillot, F.; Sarigiannidou, E.; Monroy, E. Charge Distribution and Vertical Electron Transport through GaN/AlN/GaN Single-Barrier Structures. *Semicond. Sci. Technol.* **2007**, *22* (2), 107–112. https://doi.org/10.1088/0268-1242/22/2/018.

(26) Simon, J.; Zhang, Z.; Goodman, K.; Xing, H.; Kosel, T.; Fay, P.; Jena, D. Polarization-Induced Zener Tunnel Junctions in Wide-Band-Gap Heterostructures. *Phys. Rev. Lett.* **2009**, *103* (2), 026801. https://doi.org/10.1103/PhysRevLett.103.026801.

(27) Himwas, C.; Hertog, M. den; Dang, L. S.; Monroy, E.; Songmuang, R. Alloy Inhomogeneity and Carrier Localization in AlGaN Sections and AlGaN/AlN Nanodisks in Nanowires with 240–350 Nm Emission. *Appl. Phys. Lett.* **2014**, *105* (24), 241908. https://doi.org/10.1063/1.4904989.

(28) Ajay, A.; Lim, C. B.; Browne, D. A.; Polaczynski, J.; Bellet-Amalric, E.; den Hertog, M. I.; Monroy, E. Intersubband Absorption in Si- and Ge-Doped GaN/AlN Heterostructures in Self-Assembled Nanowire and 2D Layers. *Phys. Status Solidi B* **2017**, *254* (8), 1600734. https://doi.org/10.1002/pssb.201600734.

(29) Sabelfeld, K. K.; Kaganer, V. M.; Limbach, F.; Dogan, P.; Brandt, O.; Geelhaar, L.; Riechert, H. Height Self-Equilibration during the Growth of Dense Nanowire Ensembles: Order Emerging from Disorder. *Appl. Phys. Lett.* **2013**, *103* (13), 133105. https://doi.org/10.1063/1.4822110.

(30) Birner, S.; Zibold, T.; Andlauer, T.; Kubis, T.; Sabathil, M.; Trellakis, A.; Vogl, P. Nextnano: General Purpose 3-D Simulations. *IEEE Trans. Electron Devices* **2007**, *54*

(9), 2137–2142. https://doi.org/10.1109/TED.2007.902871.
(31) Kandaswamy, P. K.; Guillot, F.; Bellet-Amalric, E.; Monroy, E.; Nevou, L.; Tchernycheva, M.; Michon, A.; Julien, F. H.; Baumann, E.; Giorgetta, F. R.; et al. GaN/AlN Short-Period Superlattices for Intersubband Optoelectronics: A Systematic Study of Their Epitaxial Growth, Design, and Performance. *J. Appl. Phys.* **2008**, *104* (9), 093501. https://doi.org/10.1063/1.3003507.

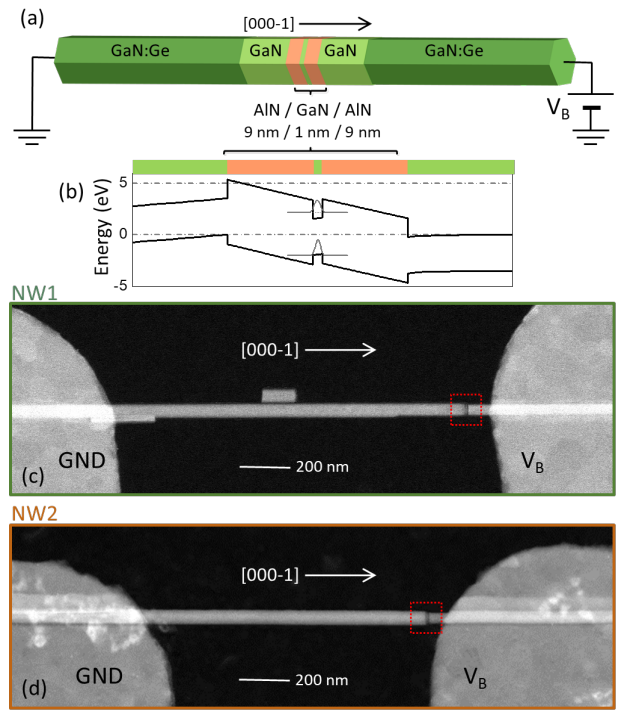

**Figure 1.** (a) Simplified schematic of the nanowire structure with the AlN/GaN/AlN. During the measurements, the nanowire stem is grounded and the bias is applied to the cap. (b) 1D nextnano³ simulations of the electric band structure of the nominal nanowire at zero bias. The squared electron and hole wavefunctions in the quantum dot are indicated. (c,d) HAADF-STEM images of (c) NW1 and (d) NW2. The location of the AlN/GaN/AlN insertion is indicated by the red squares.

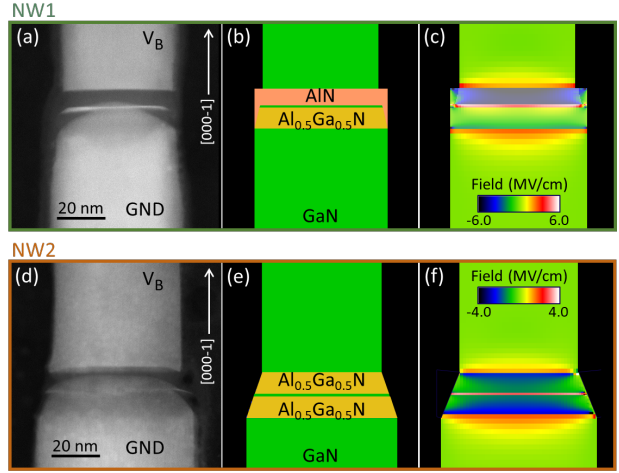

**Figure 2.** (a) HAADF-STEM micrograph of the heterostructure of NW1. (b) Material parameter definition for the 3D modeling of NW1. (c) Component of the electric field along the nanowire growth axis for NW1. The expected electric field in the GaN quantum dot is around 4.8 MV/cm. (d) HAADF-STEM micrograph of the heterostructure of NW2. (e) Material parameter definition for the 3D modeling of NW2. (f) Component of the electric field along the nanowire growth axis for NW2. The expected electric field in the dot is around 3 MV/cm.

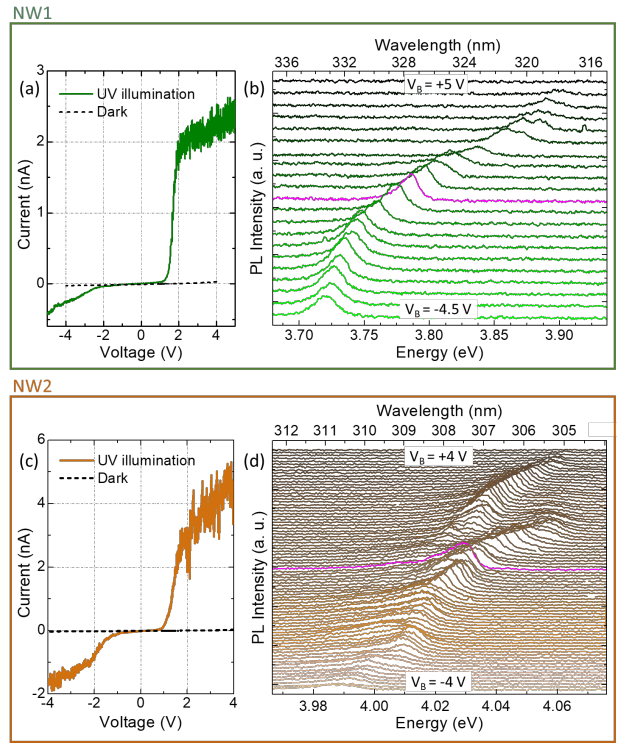

**Figure 3.** (a) I-V curve of NW1 in the dark and under ultraviolet illumination. (b) $\mu$PL spectra obtained applying from −4.5 V to +5 V bias on NW1. The zero bias measurement is indicated in magenta. The spectra are given without normalization and shifted vertically for clarity. (c) I-V curve of NW2 in the dark and under ultraviolet illumination. (d) $\mu$PL spectra obtained applying from −4 V to +4 V bias on NW2. The zero bias measurement is indicated in magenta. The spectra are given without normalization and shifted vertically for clarity.

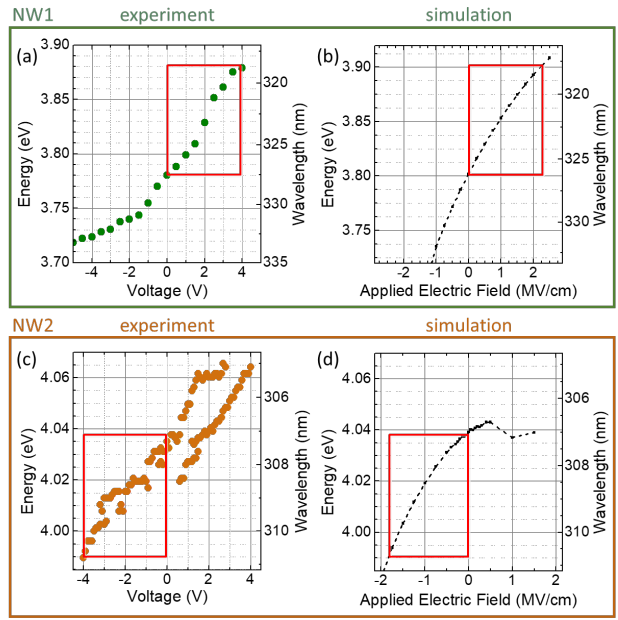

**Figure 4.** (a) µPL peak energy vs. applied bias in NW1. A shift of 100 meV is observed when varying the bias voltage from 0 to 4 V. (b) Calculation of the electron-hole transition in the quantum dot in NW1 as a function of the applied electric field. (c) µPL peak energy vs. applied bias in NW2. A shift of 48 meV is observed when varying the bias voltage from 0 to -4 V. A second peak attributed to a charged exciton state appears under positive bias (d) Calculation of the electron-hole transition in the quantum dot in NW2 as a function of the applied electric field. Red squares outline regions where the experimental results and the simulations match the best.

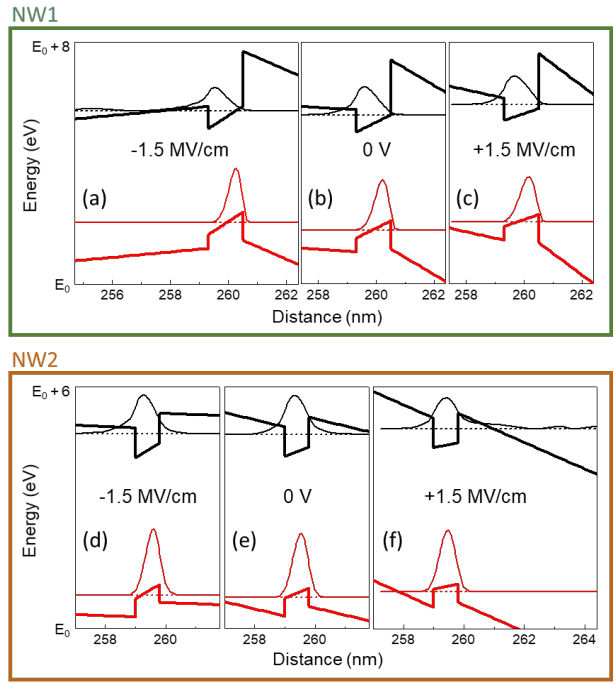

**Figure 5.** Calculation of the conduction (black) and valence (red) band around the GaN quantum dot, ground electron and hole levels in the dot, and their squared wavefunctions, for (a-c) NW1 and (d-f) NW2. The applied electric field is (a,d) -1.5 MV/cm, (b,e) 0 MV/cm, and (c,f) +1.5 MV/cm.